# Industrial graphene coating of low-voltage copper wires for power distribution

Neeraj Mishra*[#1,2], Ylea Vlamidis*[1,2], Leonardo Martini[1,2], Arianna Lanza[1], Alex Jouvray[3], Marco La Sala[4], Mauro Gemmi[1], Vaidotas Mišeikis[1,2], Matthew Perry[3], Kenneth B.K. Teo[3], Stiven Forti[1], Camilla Coletti[1,2,#]

*Equal contribution
#corresponding authors: neeraj.mishra@iit.it, camilla.coletti@iit.it

[1]Center for Nanotechnology Innovation@NEST, Istituto Italiano di Tecnologia, Piazza San Silvestro, 12−56126 Pisa, Italy
[2]Graphene Labs, Istituto Italiano di Tecnologia, Via Morego 30, 16163 Genova, Italy
[3]AIXTRON LTD, Anderson Road, Swavesey, Cambridge CB24 4FQ, United Kingdom
[4]Baldassari Cavi, Viale Europa 118/120, 55013 Capannori (Lucca), Italy

## Abstract

Copper (Cu) is the electrical conductor of choice in many categories of electrical wiring, with household and building installations being the major market of this metal. This work demonstrates the coating of Cu wires – with diameters relevant for low voltage (LV) applications – with graphene. The chemical vapor deposition (CVD) coating process is rapid, safe, scalable and industrially compatible. Graphene-coated Cu wires display oxidation resistance and increased electrical conductivity (up to 1% immediately after coating and up to 3% after 24 months), allowing for wire diameter reduction and thus significant savings in wire production costs. Combined spectroscopic and diffraction analysis indicate that the conductivity increase is due to a change in Cu crystallinity, induced by the coating process conditions, while electrical testing of aged wires shows that graphene plays a major role in maintaining improved electrical performances over long periods of time. Finally, graphene coating of Cu wires using an ambient pressure roll-to-roll (R2R) CVD reactor is demonstrated. This enables the in-line production of graphene-coated metallic wires as required for industrial scale-up.

## 1. Introduction

The applicative realm of copper (Cu) is enormous: it is utilized for indoor and outdoor constructions, tools, machinery, and portable devices, both for civil and industrial uses. In particular copper wires are commonly employed in industrial wiring and interconnection technology [1], owing to their excellent advantages of thermal and electrical conductivity, low cost, and good mechanical properties such as ductility. However, copper wires are easily oxidized when exposed to air even at room temperature, making their electrical performances waning through time in low voltage (LV) applications and limiting their use in high-power electronic devices and integrated circuits [2,3]. Therefore, nowadays research efforts are devoted to develop effective processes and coatings to prevent oxidation and corrosion of Cu wires thus preserving the electrical properties of the pristine metal over time [4,5]. Graphene represents an interesting candidate material for ultrathin coating of Cu wires thanks to its extraordinarily high electrical and thermal conductivity, flexibility, strength, and chemical inertness [6–8]. Moreover, graphene's tight structure is impermeable to gases and liquids [9,10], leading to remarkable performances as protective barrier in comparison to other thin-film materials [11]. A number of papers have discussed the potential of graphene as a multifunctional coating for Cu, hindering oxidation and chemical etching, [12–14] while increasing electrical and thermal conductivity [15,16]. Concerning Cu wires, it has been demonstrated that multilayer graphene can be synthesized on cylindrical conductors by chemical vapor deposition (CVD) [12], providing anti-oxidation protection up to 350 °C and corrosion inhibition in ammonium persulfate solution[17,18]. Furthermore, graphene-coated (Gr-coated) Cu wires with diameter $\cong$ 0.1 mm displayed improved surface heat dissipation, electrical conductivity and thermal stability up to 450 °C [18,19]. However, to date, no work has demonstrated an industrially compatible approach for this process, *i.e.*, graphene coating of cables performed complying with the following requirements: (i) process gases below the lower explosive limit (LEL); (ii) temperatures below 1000 °C; (iii) ambient pressure. Addressing all these points would pave the way to the implementation of an in-line CVD Gr-coating of metallic wires with an ambient pressure roll-to-roll (R2R) CVD system, and is the object of our work. Moreover, we address the coating of LV wires, which are expected to generate the highest revenue in the copper wire and cable market in the timeframe 2021 – 2030[20].

Indeed, this work demonstrates that wires with diameters technologically relevant for LV applications (*i.e.*, 1.37 and 1.74 mm) can be effectively coated with graphene in a scalable and industrially compatible process: at temperature below 1000 °C, ambient pressure, and with rapid growth in a non-explosive atmosphere. The Gr-coated wires, characterized by electrical measurements and microscopic and spectroscopic techniques, display higher electrical conductivity and improved aesthetics with respect to uncoated wires, allowing for significant saving in the Cu market. Aging of Gr-coated wires is investigated and in-depth studies to correlate the electrical performances and the microstructural changes after the coating are provided. It is found that the graphene growth process induces an improvement in the crystallinity of Cu wires which is responsible for the augmented conductivity and that graphene coating plays a major role in maintaining this improved conductivity over time. Finally, the production of Gr-coated Cu wires with an open-end R2R pilot CVD system is demonstrated. This is required to implement in-line graphene coating of metallic wires in an industrial setting.

## 2. Materials and Methods

### 2.1 Sample preparation

Industrial LV Cu wires were provided by Baldassari Cavi, a manufacturer of Cu wires operating in the European market.[21] Rigid Cu wires with diameters of 1.37 mm and 1.74 mm were used and these are technologically relevant for LV applications especially in Northern Europe. For each wire diameter size, three sets of samples were studied: (i) pristine Cu wires (reference); (ii) Gr-coated Cu wires: annealed at high temperature (*i.e.*, to increase Cu grain size, reduce surface roughness and defects density [16]) and then subjected to the graphene growth process (iii) only annealed Cu wires (*i.e.*, subjected to the same annealing used to process wires at point (ii)).

To facilitate the industrial translation of the process, the wires were used as-received for all tests carried out (i.e. no additional pretreatment, such as etching or chemical cleaning was done before processing the samples). Copper wires 2.5 meters in length were taken from the bundle and arranged in a coil shape, in such a way that the coil could fit inside a 4-inch BM Pro AIXTRON CVD reactor where wire treatment was performed [22] (see Figure S1 in the Supplementary Information for the reactor setup). The coils were bound with thin, flexible copper wires (diameter 250 µm) to prevent them from unwrapping and thus touching the inner walls of the reactors during heating and cooling steps. Before the growth process, argon (Ar) gas was flushed to purge the reactor from air.

The optimized process was carried out between 900–1000°C, under an Ar pressure of 750–800 mbar. Details of the temperature profile and conditions employed in the CVD process are reported in Figure S2. Cu wires that were subjected only to thermal annealing were placed in the reactor and heated at 980 °C for 10 minutes in a flow of Ar at 2000 sccm. In the case of Gr-coated samples, after the initial annealing of 10 minutes, 2 sccm of methane ($CH_4$) were flushed for 5 s in hydrogen ($H_2$) and Ar (20 and 2000 sccm, respectively) at 980 °C to grow graphene. After either annealing or growth, the samples were cooled down in Ar flow. Processed copper wires were taken out from the reactor once the temperature reached 120 °C. The process conditions were transferred from the batch reactor to the R2R CVD system.

## 2.2 Sample characterization

Raman spectroscopy was performed to assess coverage, quality, and number of graphene layers. A Renishaw InVia system was used, equipped with a 473 nm blue laser, 100× objective and 1800 l/mm grating. Spatially resolved Raman maps were obtained with 1 µm step while irradiating the samples with 22.3 mJ µm$^{-2}$. The number of graphene layers was determined by analyzing the $I_{2D}/I_G$ ratio, where $I_{2D}$ and $I_G$ are the intensities of the 2D and G peaks in the Raman spectrum, respectively [23].

The surface morphologies of the as-grown and aged wires were investigated by scanning electron microscopy (SEM, ZEISS Merlin), operated in Inlens Signal mode with a 5 kV voltage and a 120 pA current. Additionally, atomic force microscopy (AFM) was performed with a Bruker Dimension Icon in standard tapping mode. Optical microscopy (Zeiss Axioscope7 equipped with Axiocam 208 color) was employed to record bright filed micrographs and assess Cu oxidation. For the X-ray diffraction (XRD) experiments, short wires (2-3 cm long segments) were processed with the same parameters described in section 2.1, varying only the process time. In particular, analysis was done for samples annealed in Ar for 5, 10 and 30 minutes and Gr-coated samples, annealed/grown for increasing times. To further investigate the effect of hydrogen on the microstructure of copper analysis was also performed on samples annealed in Ar and then processed in $H_2$ flow, employing the same ratio of $H_2$/Ar as in the growth process. XRD measurements were carried out on a STOE Stadi P diffractometer equipped with Cu-K$\alpha_1$ radiation ($\lambda$ = 1.5406 Å) and a Ge (111) Johansson monochromator from STOE & Cie. The samples (2–3 cm long wire segments) were mounted on a goniometer head and optically aligned in the diffractometer's center of rotation and kept under spinning during the measurements. The diffracted intensities as a function of the scattering angle, 2θ, were acquired in the range 84–127 ° with a MYTHEN2 1 K detector from Dectris. The profiles of the Cu (311), (222) and (400) peaks were analyzed individually, by fitting each of them with a Pseudo-Voigt function. The dependence of the peak full-width-at-half-maximum (FWHM) on the sample treatment conditions was qualitatively evaluated for each of the three peak families and for each of the two wire thicknesses. Due to the bulk nature of the samples and the peculiar experimental geometry, the position and the intensity of the peaks are not directly providing any crystallographic information, but the variation of the normalized FWHM with respect to the pristine material is indicative of variations in crystallite strain.[24]

Electrical resistivity measurements on the Cu samples were performed in DC configuration, using a Keithley 2450 source meter. In order to minimize measurement error, we performed all the electrical measurements in a 4-probe configuration. Such a setting reduces the variation of the contact resistance between the probe and the measurement clamp due to surface oxidation. Each set of samples were tested by putting the source probe at the end of the wires, while the sensing probes were put at exactly 2 meters apart, through a couple of crocodile clips. The resistance was then measured by applying a constant DC current of 10 mA and measuring the voltage drop between the two sensing tips. All measurements were performed at room temperature. Conductivity was calculated as follows: $\sigma = L\, R^{-1}\, A^{-1}$, where R is the resistance of the cable, A is its cross-sectional area defined by design, and L is the distance between the sensing probes (2 meters for our measurements).

Also, additional electrical resistivity measurements were performed at industrial premises (*i.e.*, Baldassari Cavi) in a commercial set-up that allows measurements at controlled temperature, see Figure S3 of the Supplementary Information. A 2.5 m long coiled copper wire was kept inside the water maintained at 20 °C and firmly fixed at both ends so that there was no bending or loops. A couple of wedge holders were used as electrical sensors, while the wire holder at the end of the line provided the source current, with the distance between them that could be precisely tuned. In this way, it was possible to measure the resistivity of pristine and graphene-grafted wires avoiding any variation arising from the copper temperature or contact resistance and reducing the uncertainty on the wire length. In addition, before the resistivity was measured, a small portion of wire was cut and measured to check possible changes in diameter before and after the process. For the electrical measurements reported the average electrical conductivity was obtained from measuring 7 samples of each type.

### 3. Results and Discussion

*3.1 Gr-coated Cu wires via industrially-compatible CVD: properties*

Preliminary studies were performed to identify the process conditions to obtain graphene-coated Cu wires while satisfying the following requirements: (i) process gases below the LEL; (ii) temperatures below 1000 °C; (iii) ambient pressure. A summary of such preliminary investigations is reported in the SI (Figure S4 and S5). Notably, and different to what reported in literature until now, the optimized process shown herein has a volume percentage of explosive gases which complies with LEL requirements. Indeed, $H_2$ and $CH_4$ were used during process were at 1% and 0.1% concentration respectively and this is significantly below the LEL (i.e., 4% and 5% for $H_2$ and $CH_4$, respectively). This allows this process to operate in safe conditions and this can be easily implemented in an industrial setting (see SI). Optimized process temperature and pressure were 980°C and 780 mbar, respectively. Notably, no wire pre-treatment was necessary, besides reactor annealing, prior to graphene growth. This is also relevant as most of the work reported to date requires some form of chemical pre-treatment, thus introducing additional complexity and significant cost to the implementation of this graphene coating process at industrial premises. Table S1 presents a comparison between this work and those present in literature with respect to sample chemical pre-treatment, process temperature and gas LEL.

Figure 1 shows the appearance of the pristine (panels a, i) and Gr-coated (panels e, m) Cu wires, arranged in coil shape. After the CVD process, the Gr-coated wires display higher malleability and appear shinier and smoother when compared to the pristine ones. The optical characterization of the wire surface, shown in Figure 1 (panels b, f, j, n), reveals no evident oxidation signs, both for the pristine and for the as-processed wires. It can be further observed that the Gr-coated wires display a smoother and more homogeneous surface if compared to the pristine counterparts, for both diameters.

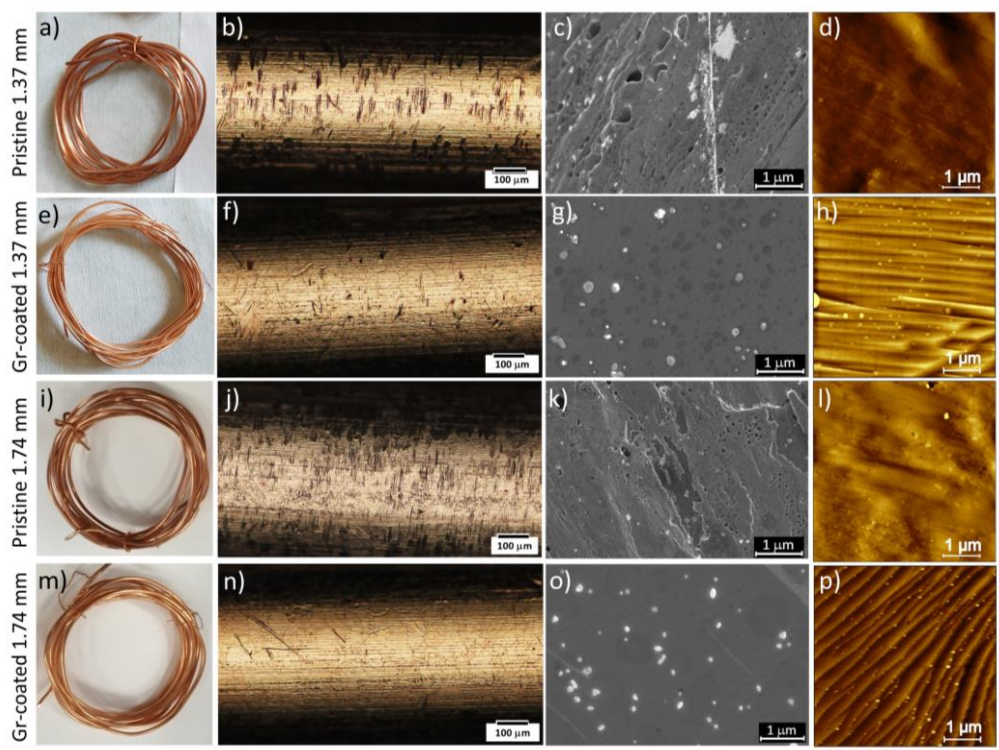

Figure 1. Morphological properties of pristine and Gr-coated Cu cables. Pictures of (a, i) pristine and (e, m) Gr-coated Cu wires arranged in coils. Optical micrographs of (b, j) pristine and (f, n) Gr-coated Cu wires. SEM micrographs of (c, k) pristine and (g, o) Gr-coated Cu wires. AFM topography images of (d, l) pristine and (h, p) Gr-coated Cu wires. All data are reported for both the 1.37 and 1.74 mm wire diameters.

To study in detail the surface morphology, SEM and AFM characterization for all the samples were performed. Figure 1 c, d and k, l report SEM and AFM micrographs of pristine Cu wires, 1.74 and 1.37 mm diameter, respectively. For both diameters, an inhomogeneous surface can be observed, with an RMS roughness of 6.5 and 5.6 nm for 1.37 and 1.74 mm (over areas of 1 µm$^2$), respectively. SEM micrographs of the Gr-coated wires display a smoother wire surface decorated with particles of ~160 nm size. AFM micrographs reveal the presence of atomic terraces, typically observed after the growth of graphene with good crystallinity and thickness homogeneity [25–28]. The RMS roughness measured from 1 µm$^2$ micrographs for the 1.37 and 1.74 mm wires is ~3.9 nm and 4.6 nm, respectively.

Raman spectroscopy confirmed the presence of graphene films on the wires of both diameters. As visible in Figure 2c sharp G (~1590 cm$^{-1}$) and 2D bands (2720 cm$^{-1}$) are visualized. The average $I_{2D}/I_G$ ratio over an area of 30x30 µm$^2$ was found to be ~0.9 and ~0.80 for 1.37 and 1.74 mm diameters, respectively (Figure 2b,e). Similarly, the average FWHM (2D) was measured as 65 and 63 cm$^{-1}$ for 1.37 and 1.74 mm diameters, respectively (Figure 2a, d). Combined analysis of FWHM (2D) and I(2D)/I(G) indicates that the average number of layers is 2-3 [29]. Furthermore, the as-grown samples exhibit a negligible D peak (see Figure 2c), indicating a negligible number of defects. No sp2-related bands are observed for pristine and annealed wires in the same wavenumber range.

The electrical conductivity of Gr-coated wires measured at controlled temperature with a commercial system at industrial premises is reported in Figure 2f as a percentage variation with respect to pristine wires. Immediately after growth, Gr-coated wires displayed an average electrical conductivity improvement of 0.98% (for 1.37 mm) and 0.74% (for 1.74 mm). It should be mentioned that already a 0.6% conductivity improvement would lead to a wire diameter reduction corresponding to a Cu cost reduction of 200€/Ton [30,31], an appealing saving for Cu cable manufacturers.

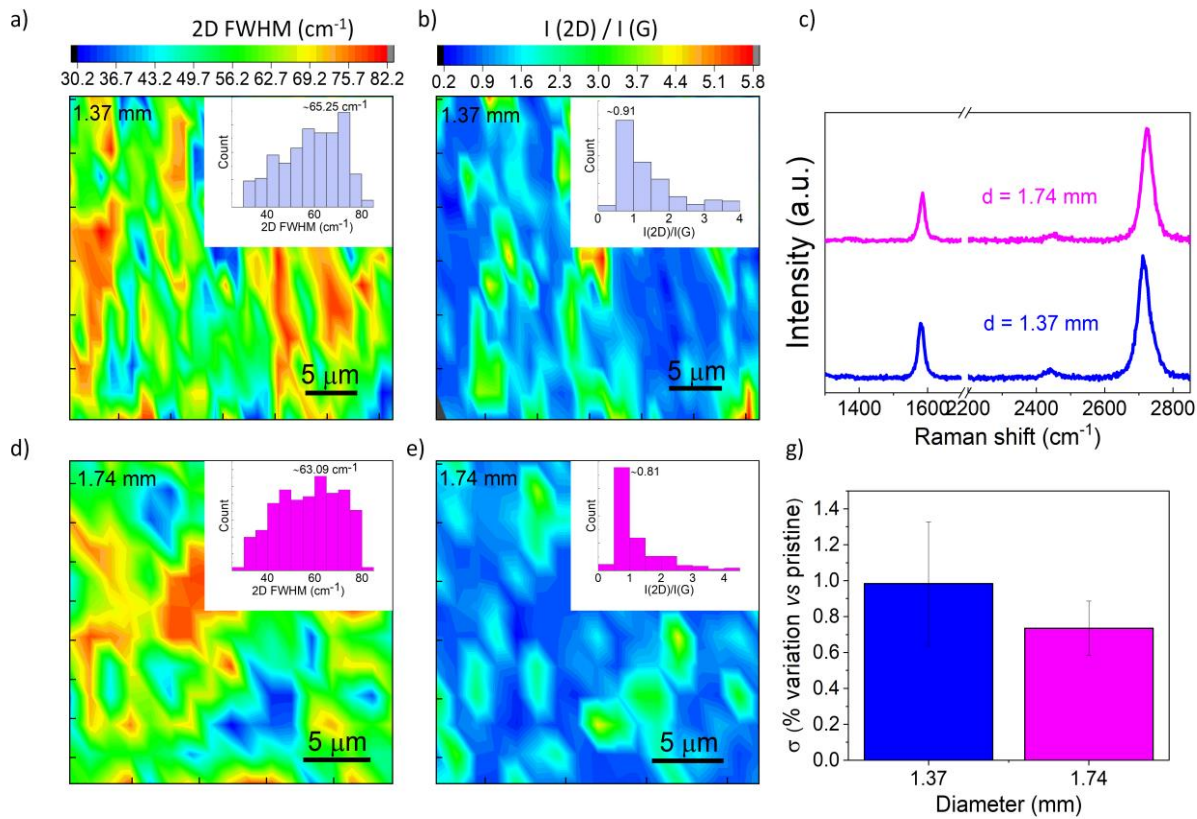

Figure 2. Spectroscopic and electrical properties of Gr-coated Cu wires. Representative Raman 30x30 µm² maps (with relative histograms in the insets) reporting the FWHM (2D) (a, d) and the intensity ratio of 2D/G bands (b, e) for both thicknesses. Representative Raman spectra for 1.37 and 1.74 mm wires (c). Electrical conductivity improvement of Gr-coated wires with respect to pristine ones (f). Error bars indicate the standard deviation of the mean (SDM).

### 3.2 Aging of Gr-coated Cu wires

To assess the performances of aged Cu wires, a complete characterization of the samples by means of optical microscopy, Raman spectroscopy and electrical measurements was carried out. The effect of aging on the chemical properties was investigated at different time points, from 6 months to 24 months after the CVD process.

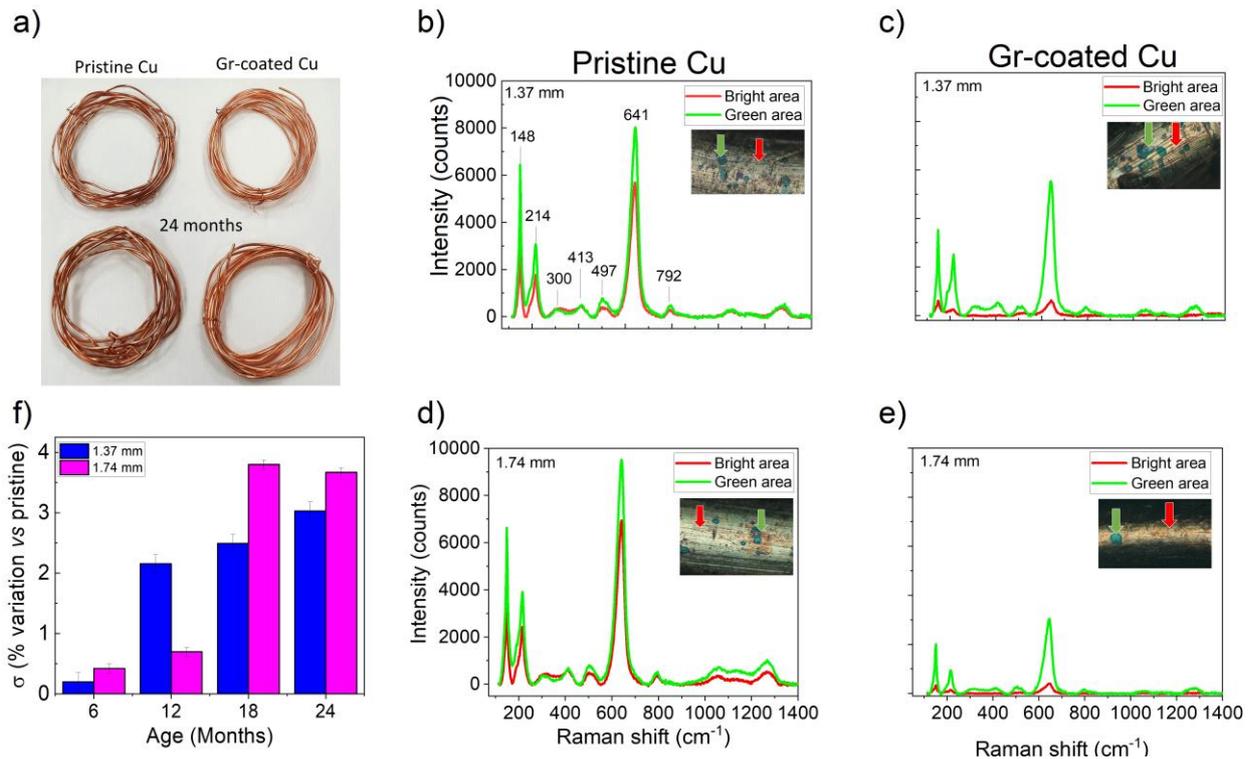

Figure 3. Characterization performed after aging of pristine and Gr-coated Cu wires. Optical image of pristine and Gr-coated wires after 24 months (a). Raman spectra taken on different areas (indicated by arrows in the insets) of pristine copper wires (b and d) and Gr-coated copper wires (c and e), for both thicknesses after 24 months Electrical conductivity improvement of 1.37 and 1.74 mm Gr-coated wires with respect to pristine wires from 6 to 24 months of aging (f).

Figure 3a shows a picture of pristine and Gr-coated Cu wires after 24 months of aging: color darkening – a clear sign of oxidation – is very evident in uncoated wires, while the Gr-coated ones maintain a shiny and new appearance. Figure S6 confirms a similar trend for both kinds of wires also at intermediate aging times. It is well known that Cu wires tend to become darker over time: this is usually thought as an indication of "poor quality" and causes significant sale returns which negatively affect Cu wire manufacturers and retailers. Graphene coating offers a viable solution to this issue, providing an aesthetic advantage.

Optical microscopy analysis might apparently contradict the resistance to oxidation of Gr-coated wires, since oxidation spots (observed as greenish colored areas) can be found on both pristine and Gr-coated wires (see insets in Figure 3 b-e). However, Raman analysis confirms that Gr-coated wires have a higher oxidation resistance. Raman spectra were recorded in the copper oxides ($Cu_xO$) range (between 150 and 800 cm$^{-1}$) [32] , on both the green and bright areas (Figure 3b-d, c-e). Remarkable differences can be observed when comparing the uncoated and Gr-coated samples. The pristine wires (Figure 3 b, d) display a very high intensity of $Cu_xO$ peaks both in the green and in the bright areas, demonstrating an extended oxidation which involves the entire surface of the wire. On the contrary, in the case of Gr-coated wires (Figure 3 c, e) $Cu_xO$ peaks with a significantly lower intensity than those found in pristine wires are measured in the green areas, with the rest of the surface having a negligible $Cu_xO$ signal. The electrical properties of the wires at room temperature were also investigated for different aging times by recording their resistance at 10 mA. The relative improvement of the electrical conductivity, calculated with respect to the pristine Cu wires at the same aging stage, is reported in Figure 3f.

The different electrical behavior for wires with and without graphene coating over 6 to 24 months is apparent. After 12 months, a gradual worsening of the electrical properties for the pristine wires was observed, while the resistivity of the Gr-coated Cu wires remained comparable to that of the as-grown ones. This result further confirms that the presence of graphene significantly reduces the deterioration of the Cu properties. Remarkably, after 24 months a relative improvement in conductivity of 3.0% and 3.6% is observed for the 1.37 and 1.74 mm Gr-coated samples, respectively. The waning of electrical conductivity in the pristine wires is associated to the oxidation of Cu discussed above [18,33].

*3.3 The effect of wire annealing and hydrogen treatment on wire microstructure*

From the analyses reported above it appears that Gr-coated wires have the following advantages with respect to pristine ones: (i) increased electrical conductivity at time zero and overtime; (ii) oxidation protection. One could argue that the increase in electrical conductivity could be induced by microstructural changes in the Cu wires due to processing conditions rather than from the presence of the graphene thin layer itself. Indeed, on millimeter-sized wires it is quite unlikely that a thin layer of graphene could have such a remarkable effect on the electrical properties, although such an explanation has been provided for thinner wires.[19] In order to assess the influence of the thermal annealing on the electrical properties of the Cu wire, the electrical conductivity of the annealed wires over time was investigated. Despite exhibiting a shiny appearance (Figure S7) and an improved conductivity at t=0, it is found that the conductivity improvement over time is lower than that measured for Gr-coated wires (see Figure 4a). Hence, the presence of graphene seems pivotal for maintaining augmented electrical performances. Indeed, Raman analyses (as well as optical imaging) of aged wires indicate that oxidation progresses in annealed cables in a similar fashion than in pristine ones (see Figure 3b and 4b). After 24 months, annealed wires appear significantly darker than the Gr-coated, and $Cu_xO$ peaks are comparable to those measured for pristine wires of the same age (Figure S8). With the aim of further investigating qualitative microstructural differences in pristine, annealed and Gr-coated wires, XRD measurements were performed. The qualitative assessment of the XRD peaks further clarifies the role of the annealing and graphene growth steps on the conductivity of copper. The histograms in Figure 4 (c, d) report the relative variation (in %) of the FWHM of each peak with respect to the pristine samples, against the total duration of the thermal treatment (the XRD patterns are shown in the SI, Figure S9). The annealed wires display narrower FWHM if compared to the pristine Cu wires and further FWHM reduction is observed upon Gr-growth. The FWHM is sensitive to the variation in microstructure (crystallinity, presence of defects and grain size) and stress-strain accumulation in the material.[34] During the annealing and recovery process, the improved atomic diffusion at high temperature enables the release of the stored strain energy of the extruded wires. The decrease of XRD peak broadening is suggestive of a decrease of crystallite strain which can explain the enhanced conductivity of Cu after annealing. The growth step implies a remarkable drop in peak width regardless the annealing time (see SI, Figure S10). This observation suggests that the conditions used for graphene-coating relieve crystallite strain much more effectively than annealing in Ar does.

The aforementioned microstructural observations, as well as the increased malleability and smoothness, confirm that the enhanced conductivity is due to a more radical and deep change in the bulk copper crystallinity occurring during the growth treatment rather than to the presence of graphene itself. Since the growth step employs a mixture of $H_2/CH_4$, the effect of hydrogen on the microstructure of copper was investigated, analyzing samples annealed in Ar and then processed in $H_2$ flow. The details of the samples and process conditions are reported in Table S2 of the SI. As demonstrated by the histograms reported in Figure 4 for a process where the standard annealing was associated to subsequent hydrogen treatment for 5 s (without concomitant graphene growth), the peak widths are comparable to those obtained after 5 s growth, or possibly even narrower in the case of 1.37 mm wires.

Additional XRD measurements were performed for different annealing and hydrogen treatment times (see Figure S10 and Table S2), which further confirm that high-temperature $H_2$ treatment positively affects Cu crystallinity.[35] In conclusion, we can identify the hydrogen gas present in the CVD process during graphene growth as the main player in the improvement of Cu crystallinity and the consequent enhanced electrical properties; however, the structural, chemical and electrical measurements performed for the different wires indicate that graphene is necessary for maintaining enhanced electrical properties over time.

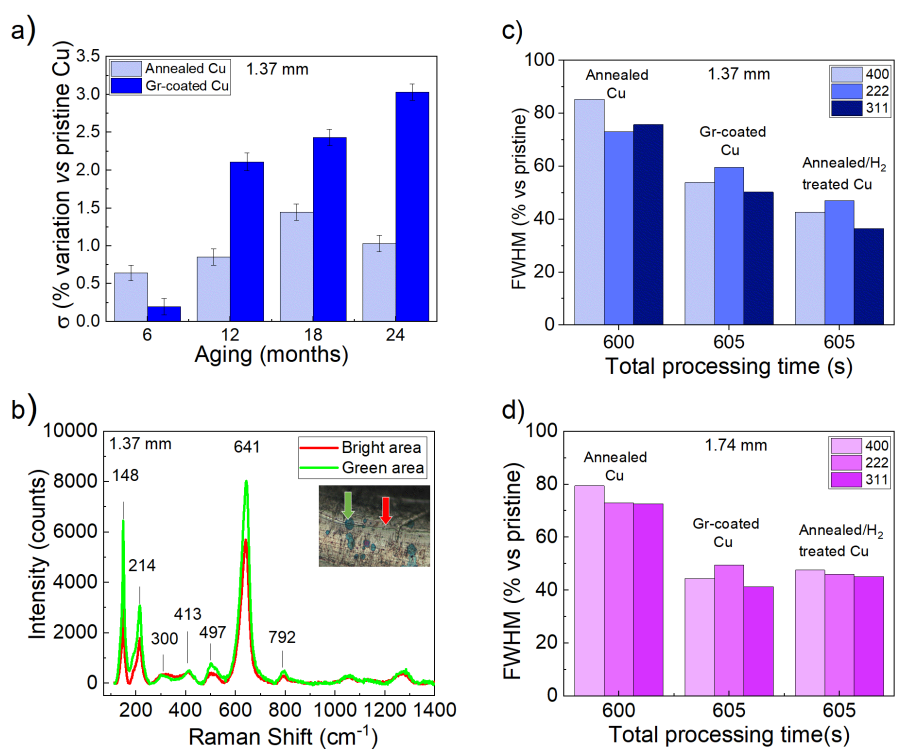

Figure 4. Electrical conductivity improvement of annealed and Gr-coated Cu with respect to pristine wires from 6 to 24 months of age (a). Raman spectra of annealed Cu taken in different areas of the wire (see inset) after 24 months of aging in $Cu_xO$ range (b). FWHM evolution of (400), (222) and (311) diffraction peaks for Cu wires which were annealed, Gr-coated and annealed plus $H_2$ treated. Both diameters were considered (c, d).

*3.4 Gr-coated Cu wires with an ambient pressure R2R CVD reactor*

The results above show that graphene coating of Cu wires is instrumental to obtain combined improved electrical conductivity over time and oxidation resistance. This makes graphene an appealing coating for LV Cu wires, while graphene coating of metals might be of interest also for other applications spanning from nautical, electrical vehicles to aerospace.[36–39] However, the scalable coating of metallic wires with graphene can be enabled only by the development of an ambient pressure R2R CVD reactor. In this work, the properties of Gr-coated Cu wires obtained with such a R2R CVD reactor developed by AIXTRON Ltd is presented. Using this in-line R2R CVD system, graphene was grown successfully onto a 1.74 mm Cu wire and the relevant characterizations are reported in Figure 5. Raman spectroscopy indicates a uniform few-layer graphene film. The FWHM (2D) averages at ~68 cm$^{-1}$, the 2D/G intensity ratio is found to be ~0.64 and the D/G intensity ratio is ~0.33 (see Figure 5a-c). XRD measurements revealed an increased Cu crystallinity, comparable to that of Gr-coated wires processed in a batch reactor (see Figure 4d and 5e). Optical and SEM characterization confirmed the presence of a uniform coating across the surface (Figure 5 f). These results indicate that the process developed in laboratory settings can be implemented in an industrial in-line system with promising results.

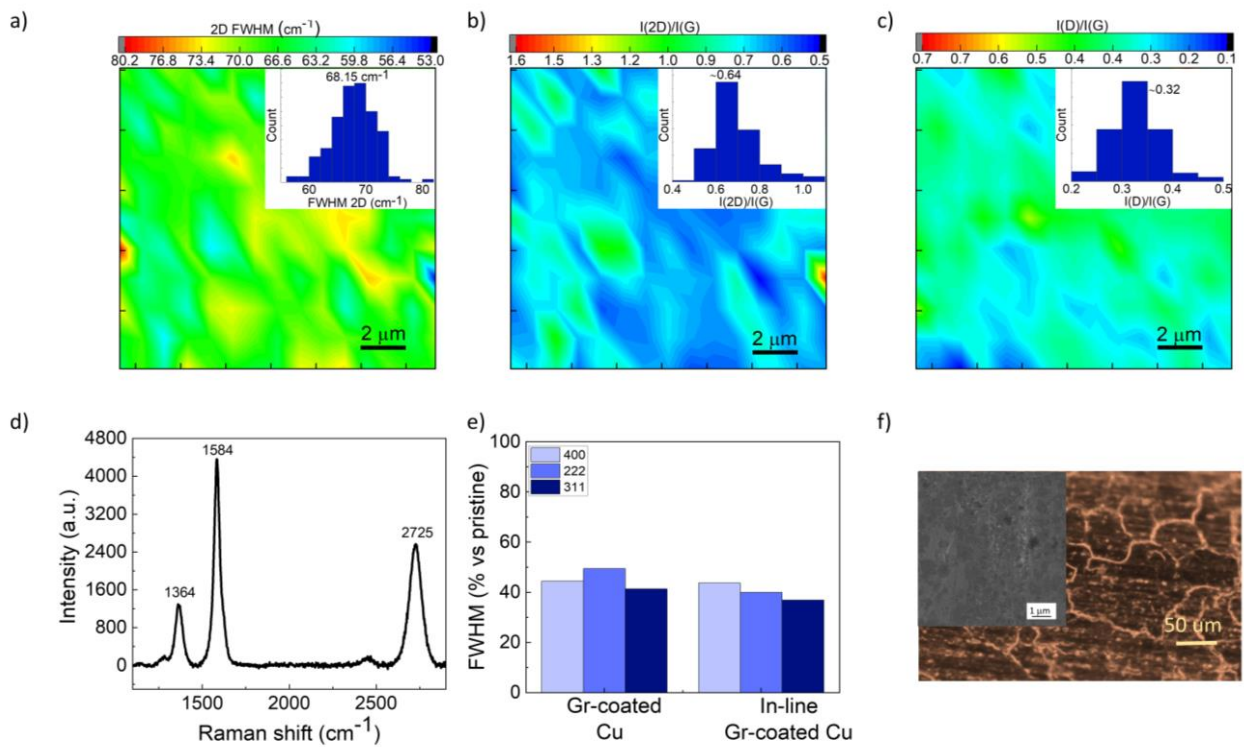

Figure 5. Characterization of Cu wires Gr-coated in a R2R prototype reactor. Representative Raman 25x25 µm$^2$ maps (with relative histograms in the insets) reporting the FWHM(2D) (a), the intensity ratio of 2D/G (b) and D/G bands (c). Representative Raman spectrum of graphene grown on 1.74 mm Cu wire (d). Comparison of FWHM evolution of (400), (222) and (311) diffraction peaks for Cu wires coated with graphene in lab and In-line reactor (e). Optical image with SEM micrograph in inset (f).

## 4. Conclusions

In conclusion, an industrially compatible process to coat LV electrical Cu wires with graphene is demonstrated. The Gr-coated wires display improved aesthetics and enhanced electrical conductivity, which is maintained for at least 24 months. Structural analyses indicate that high temperature annealing causes an increase in Cu crystallinity, that is further augmented thanks to the presence of $H_2$ during the graphene growth process. Combined electrical and chemical characterization show that graphene efficiently acts as a barrier against oxidation which is beneficial to preserve the improved conductivity in time. The reported increase in electrical conductivity potentially allows for a reduction in wire diameter with consequent savings in wire production costs. Indeed, the last years have seen an ever-increasing trend in the market-price of raw Cu that doubled from 2020 to 2022. The global copper wire and cable market was valued at 147.53 billion € in 2020, and is projected to reach 252.49 billion € by 2030.[20] Implementing a technology allowing for an increase in conductivity of about 1% could affect the market of LV wires with significant savings, estimated to be about 100 €/ton, which in the European market translates to over 50 Million €/year. Also, the oxidation protection offered by the graphene coating would have an additional positive aesthetic effect on the market: Cu wires have a tendency to become darker over time, which is usually thought as an indication of "poor quality" from a customer's perspective and this has a negative impact on sales and revenue for Cu wire manufacturers and retailers. By preventing aesthetically driven sales return, Gr-coated wires would have a longer shelf life. Building on such promising premises, we demonstrate that the developed process can be adopted in a R2R CVD reactor for in-line coating of metallic wires. For a large enterprise manufacturing LV Cu wires, it is estimated that implementing such in-line coating technology within production premises would ensure a quick break-even (i.e., ~ 2 years) with the initial investment. Furthermore, the extension of such coating technology also to flexible Cu cables and wires adopted in automotive and aerospace industries could lead to a significant decrease in the overall amount of Cu weight, leading to lighter vehicles with a lower impact on the environment. Ultimately, the demonstration of Gr-coated wires adopting a pilot-line-compatible R2R system builds a road towards the realistic translation of the graphene coating technology in industrial settings.

## Author Contributions

N.M. and Y.V. - conceptualization, methodology, investigation, writing – original draft, visualization, L. M. - investigation, writing – reviewing and editing, A.L. and M.G. - investigation, writing – reviewing and editing, V.M. - investigation, writing – reviewing and editing, M.P., A.J., and B.K.T. investigation, writing – reviewing and editing, M.L.S - writing – reviewing and editing, S. F. - investigation, writing – reviewing and editing, C.C. - conceptualization, investigation, writing –reviewing and editing, visualization, supervision.

## Acknowledgments

The research leading to these results has received funding from the European Union's Horizon 2020 research and innovation program under grant agreements no. 785219-Graphene Core2 and 881603-Graphene Core3. Neeraj Mishra and Ylea Vlamidis contributed equally to this work.

# Supplementary Information

# Industrial Graphene Coating of Low-Voltage Copper Wires for Power Distribution

Neeraj Mishra*[1,2], Ylea Vlamidis*[1,2], Leonardo Martini[1,2], Arianna Lanza[1], Alex Jouvray[3], Marco La Sala[4], Mauro Gemmi[1], Vaidotas Mišeikis[1,2], Matthew Perry[3], Kenneth B.K. Teo[3], Stiven Forti[1], Camilla Coletti[1,2,*]

*Equal contribution

#corresponding Author: neeraj.mishra@iit.it ; camilla.coletti@iit.it

[1]Center for Nanotechnology Innovation@NEST, Istituto Italiano di Tecnologia, Piazza San Silvestro, 12−56126 Pisa, Italy

[2]Graphene Labs, Istituto Italiano di Tecnologia, Via Morego 30, 16163 Genova, Italy

[3]AIXTRON LTD, Anderson Road, Swavesey, Cambridge CB24 4FQ, United Kingdom

[4]Baldassari Cavi, Viale Europa 118/120, 55013 Capannori (Lucca), Italy

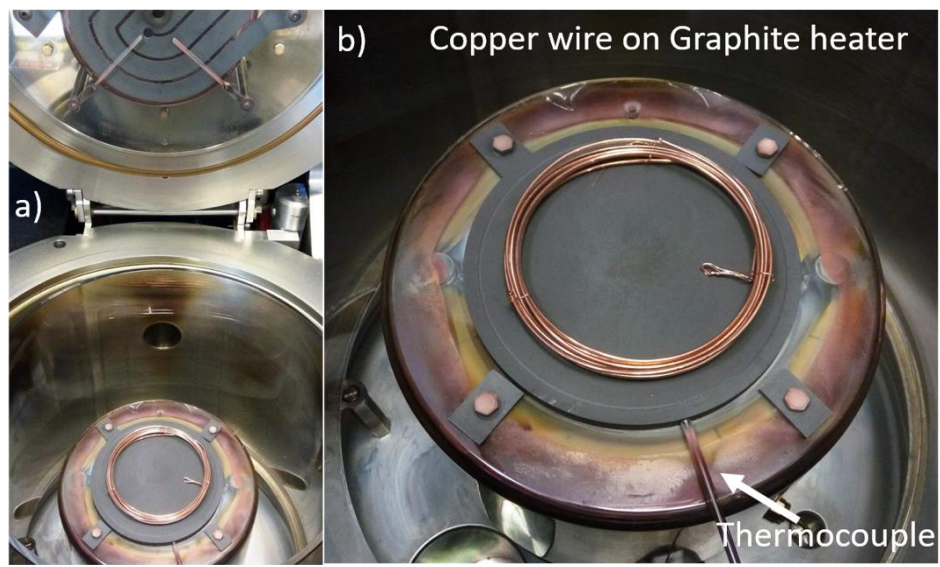

Figure S1. Copper (Cu) coil placement within the 4-inch cold wall AIXTRON BM reactor. Inside view of the reactor with lid (a), featuring a top heater and inlet for the gases, visible. Zoom-in image showing the Cu coil on top of the graphite heater (b). The thermocouple is indicated by an arrow.

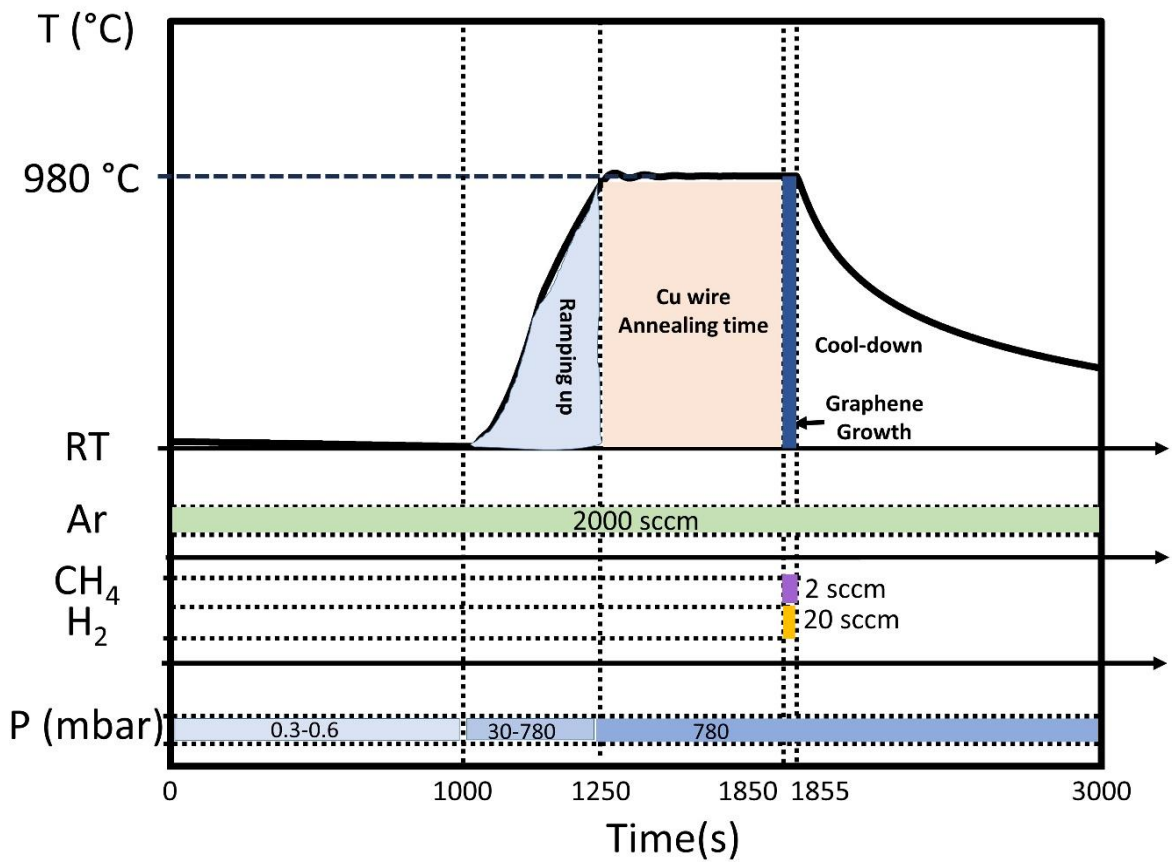

Figure S2. Schematic diagram showing the temperature profile and conditions employed in the CVD process.

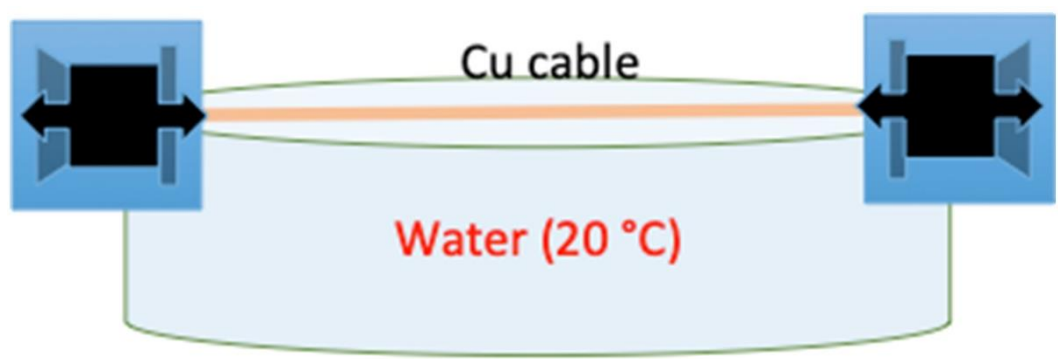

Figure S3. Standard set-up for resistivity measurements at 20 °C.

**CVD growth process optimization**

Graphene was grown on copper wire of both diameters (1.37 mm and 1.78 mm) at different temperatures, pressures, gas fluxes, annealing and growth times. Concerning pressure, 780 mbar were found to be an ideal compromise to obtain graphene growth at nearly atmospheric pressure. Argon gas was used during the annealing step to increase Cu grain size.[1] Annealing times adopted during the preliminary experiments were 0, 5, 10 and 30 min (Figure S4) at different temperatures (900, 930, 950, and 980 °C). The annealing step was found to improve surface morphology and promote Cu terrace formation. An annealing time of 10 min at 980 °C was found to be a good compromise, yielding well-defined terraces when compared to 0 s and 5 min annealing, while avoiding lengthy processing times (FigureS4). Adopted growth temperatures were 900, 930, 950, and 980 °C (FigureS5). Growth temperatures lower than 950 °C yielded to the formation of amorphous carbon for all the growth times adopted (0 s, 1 s, 5 s, 10 s, 5 min, 10 min, 30 min) as found by Raman spectroscopy analyses (not reported). A growth temperature of 980 °C for 5 s was found to yield continuous graphene. The SEM micrographs reported in Figure S5 show a significant improvement of the Cu surface morphology, instrumental for the growth of good quality graphene, with increasing growth temperature.

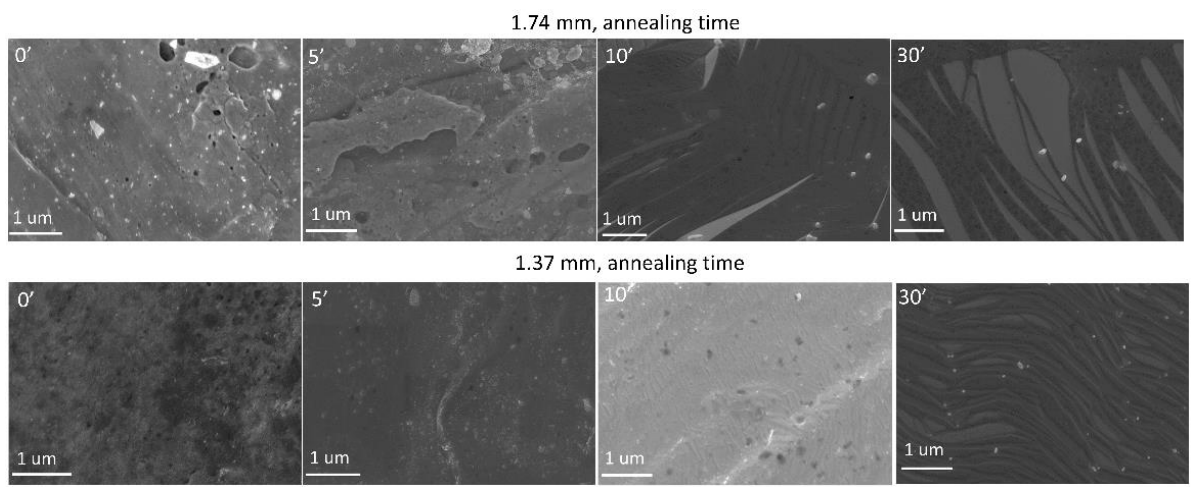

Figure S4. SEM images of copper wire of both diameters (i.e., 1.74 mm and 1.37 mm) annealed at 980°C in Ar atmosphere for 0 min, 5 min, 10 min and 30 min.

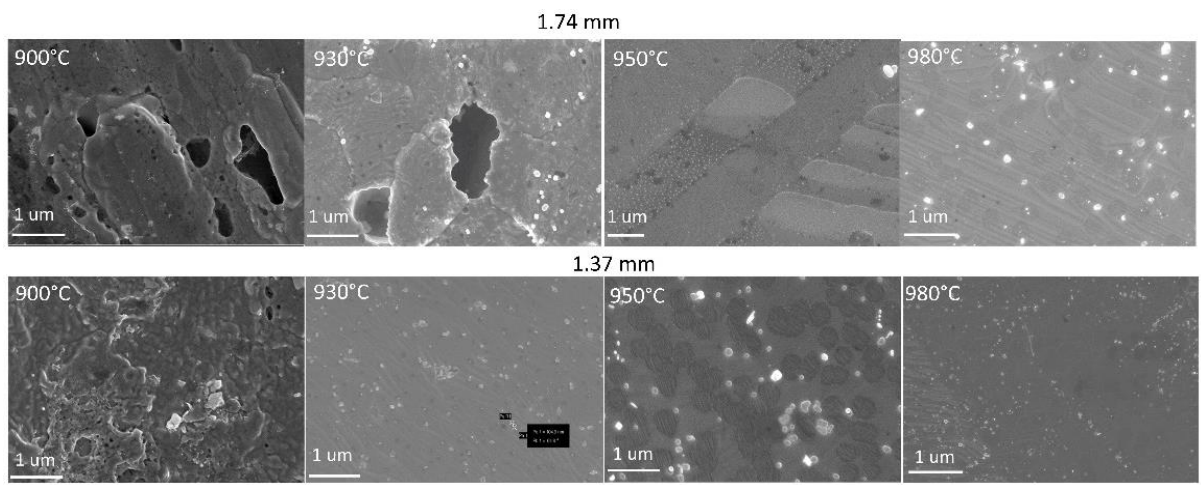

Figure S5. SEM images of copper wires of both diameters (i.e., 1.74 mm and 1.37 mm) after graphene growth for 5 s at temperatures of 900, 930, 950 and 980°C.

**Lower and upper explosive limits**

From an industrial point of view, to comply with safety requirements, it is important to maintain the formation of explosive gas mixtures below well-defined limits. Keeping this in mind, we optimized the flow of $H_2$ and $CH_4$ to be within the lower explosive limit (LEL) that is the lowest concentration of an explosive gas which will burn or explode if ignited. The LEL is determined empirically for each pure chemical and air mixture at a given temperature. If more than one chemical is dispersed in the air, as is normally the case, then Le Chatelier's mixing rule can be applied to get the cumulative LEL for the mixture.[2,3] Concentrations lower than the lower explosive limit are 'too lean' to burn; those above the Upper Explosive Limit (UEL) are too rich to burn. The amount of gas present is specified as a percentage (%) of LEL. Zero percent Lower Explosive Limit (0% LEL) denotes a combustible gas-free atmosphere.[2,3] The LEL values for $H_2$ and $CH_4$ are 4% and 5%, respectively.[3,4] Since the % volume used in our reactor is 1% and 0.1% for $H_2$ and $CH_4$, respectively, we operate in safe conditions, employing concentrations lower than the explosive limit.

Table S1. Comparison of process parameters employed in related works.

| Reference | Pre-treatment | Process T (°C) | Gas flow (sccm) | | | Explosive gases (% vol.) | |
|---|---|---|---|---|---|---|---|
| | | | Ar | $H_2$ | C source | $H_2$ | C source |
| **Lee et al.[5]** | Ammonium persulfate | 1050 | | 10 | 0.1 ($CH_4$) | - | - |
| **Jang et al.[6]** | Ammonium persulfate | 1000 | 500 | 100 | 2 ($CH_4$ or acetylene) | 16.6 (> LEL) | 0.3 (< LEL) |
| **Datta et al.[7]** | Electropolishing | 1000 | 100 | 500 | 6 ($CH_4$) | 82.50 (> UEL) | 1 (<LEL) |
| **Kashani et al.[8]** | / | 975-1000 | 1500 | 100 | 2-20 (Benzene) | 6.3 (> LEL) | 0.1-1.2 (<LEL) |
| **This work** | / | 980 | 2000 | 20 | 2 ($CH_4$) | 1 (<LEL) | 0.1 (<LEL) |

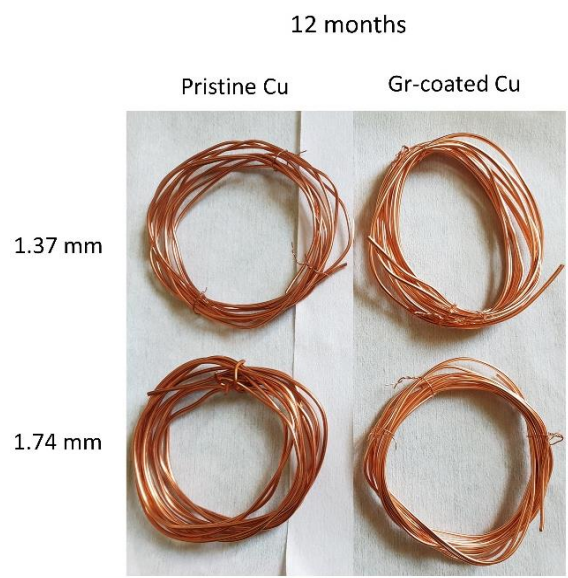

Figure S6. Optical images of pristine and Gr-coated Cu wires at 12 months.

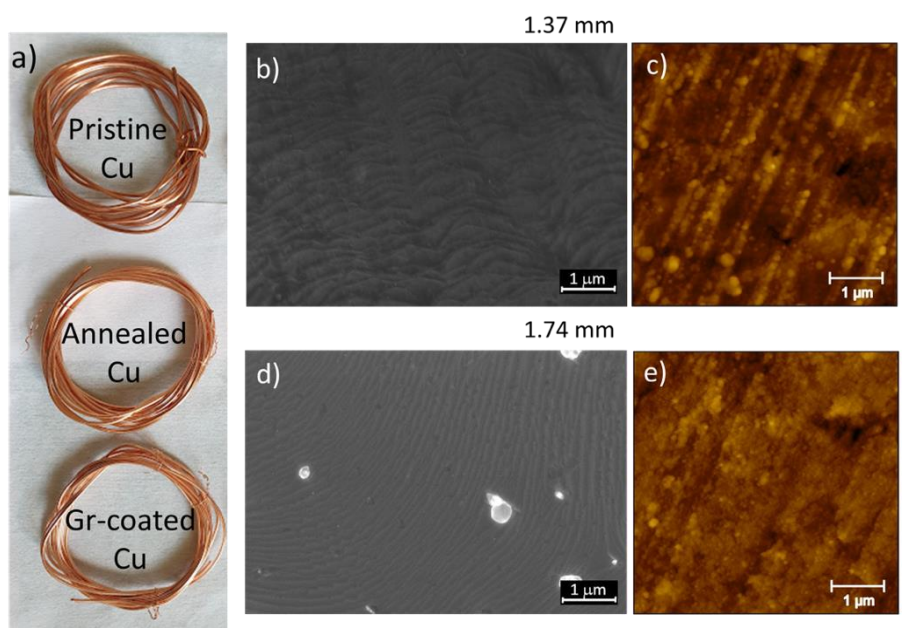

Figure S7. Optical (a), SEM (b, d) and AFM images (c, e) of annealed Cu wires. Both diameters are reported.

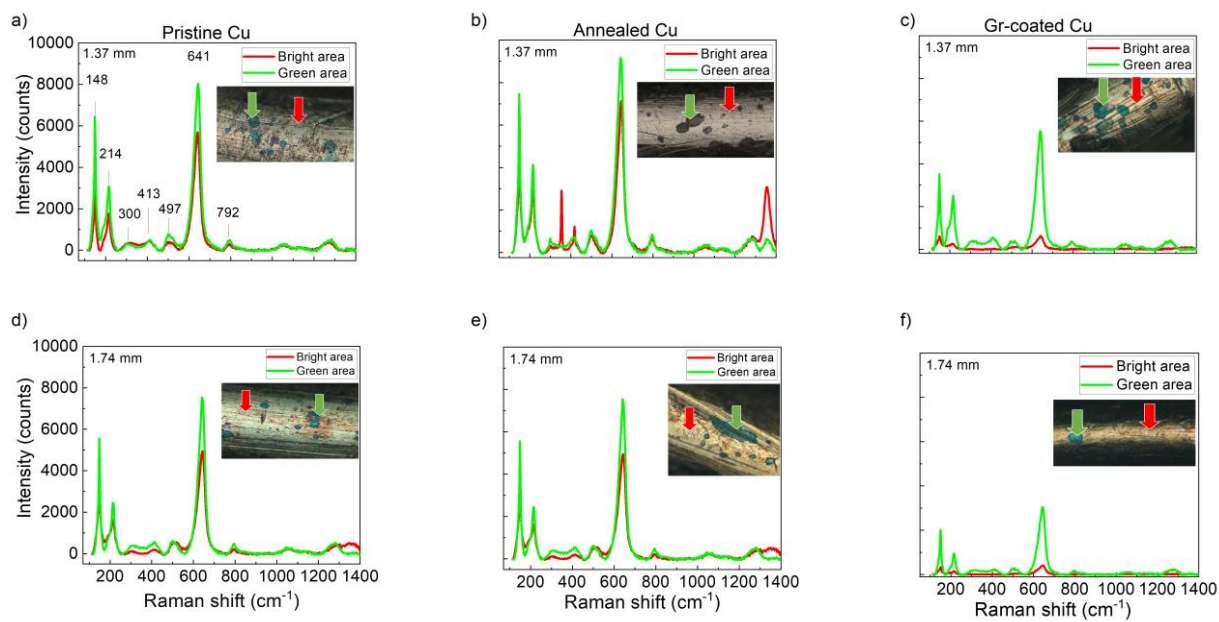

Figure S8. Comparison between Raman spectra recorded in the green areas and in the brighter areas for the pristine Cu (a, d), the annealed Cu (b, e) and grown samples (c, f) of both diameters after 24 months.

Table S2. Samples analyzed by XRD and related parameters and process conditions.

| Sample | Annealing | | Growth/H$_2$ atmosphere | | | | Total time (s) |
|---|---|---|---|---|---|---|---|
| | Ar flow (sccm) | Time (min) | Gas flow (sccm) | | | Time (s) | |
| | | | CH$_4$ | H$_2$ | Ar | | |
| pristine | --- | --- | --- | --- | --- | --- | 0 |
| 0'/1" H$_2$ | --- | --- | --- | 20 | 2000 | 1 | 1 |
| 0'/5" | --- | --- | 2 | 20 | 2000 | 5 | 5 |
| 0'/5" H$_2$ | --- | --- | --- | 20 | 2000 | 5 | 5 |
| 1'/5" | 2000 | 1 | 2 | 20 | 2000 | 5 | 65 |
| 5'/0" | 2000 | 5 | --- | --- | --- | --- | 300 |
| 5'/5" | 2000 | 5 | 2 | 20 | 2000 | 5 | 305 |
| 10'/0" | 2000 | 10 | --- | --- | --- | --- | 600 |
| 10'/1" | 2000 | 10 | 2 | 20 | 2000 | 1 | 601 |
| 10'/5" | 2000 | 10 | 2 | 20 | 2000 | 5 | 605 |
| 10'/5" H$_2$ | 2000 | 10 | --- | 20 | 2000 | 5 | 605 |
| 10'/30" | 2000 | 10 | 2 | 20 | 2000 | 30 | 630 |
| 30'/0" | 2000 | 30 | --- | --- | --- | --- | 1800 |

**XRD experiments**

As shown in Figure S9, the FWHM decreases sharply for just 1 s growth, and becomes extremely narrow for 5 s growth, while by further prolonging the growth duration (30 s) no appreciable differences are observed compared to 5 s. Considering samples annealed for 10 minutes and grown for 5 s the observed FWHM are roughly half with respect to those of the pristine samples. All peak families of for both wire thicknesses are aligned with this trend.

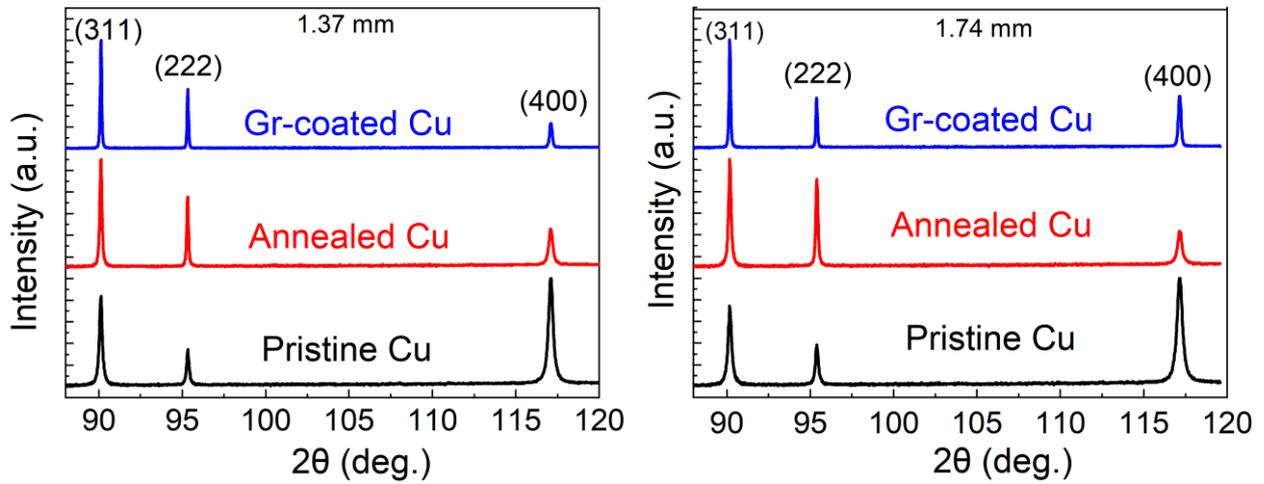

Figure S9. Representative XRD patterns of pristine and processed Cu wires, normalized to the highest intensity value (a: 1.37 mm and b: 1.74 mm size).

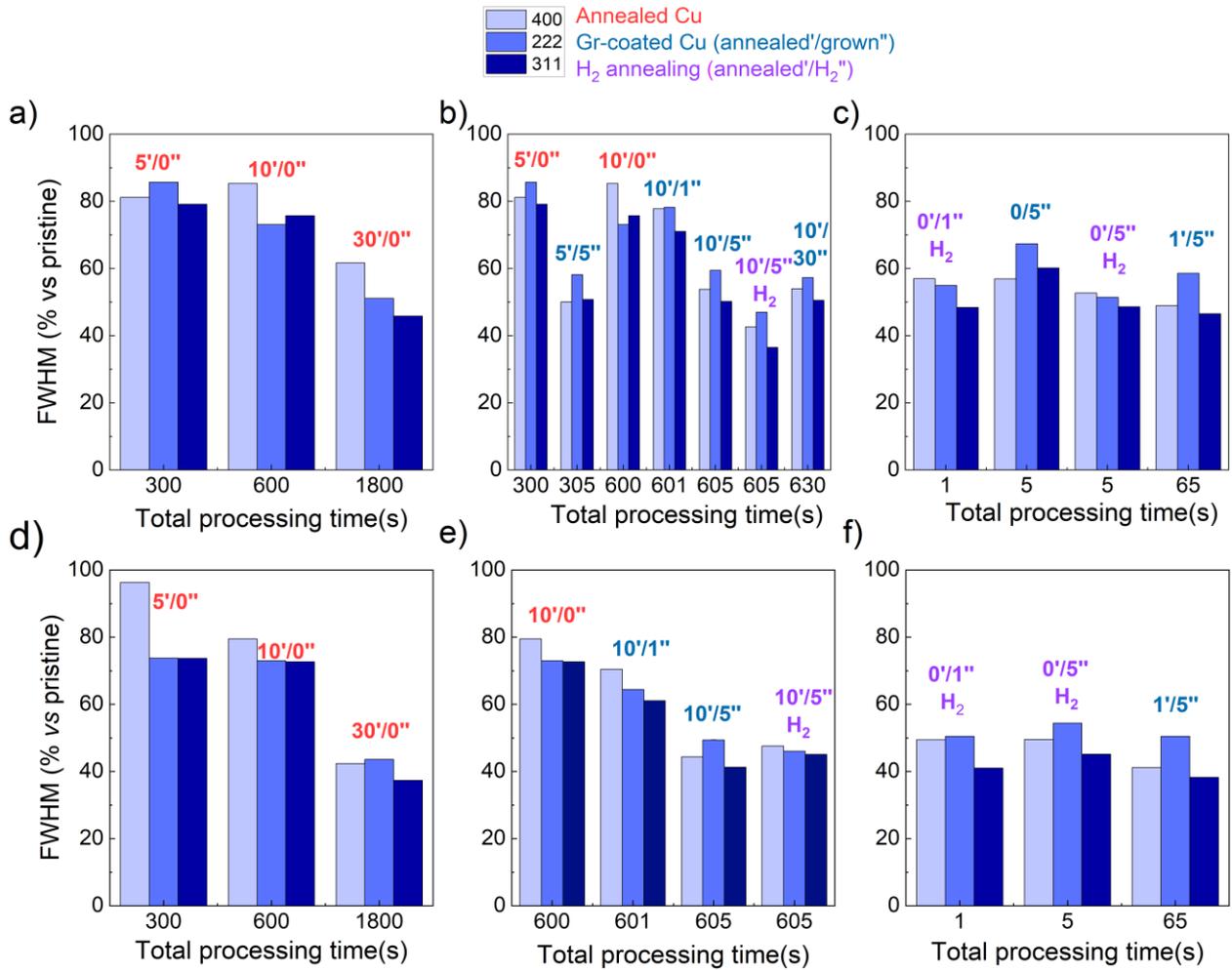

Figure S10. FWHM evolution of (400), (222) and (311) diffraction peaks for Cu submitted to different processing times: annealed samples (a, d), comparison between Gr-coated samples (annealed/grown) and wires annealed/processed in H$_2$ atmosphere (b, e), comparison between samples processed in hydrogen atmosphere and grown samples (c, f). Notation: annealed'/growth'' or annealed'/H$_2$''. Both diameters were considered (a-c: 1.37 mm, d-f: 1.74 mm).

**Graphical abstract**

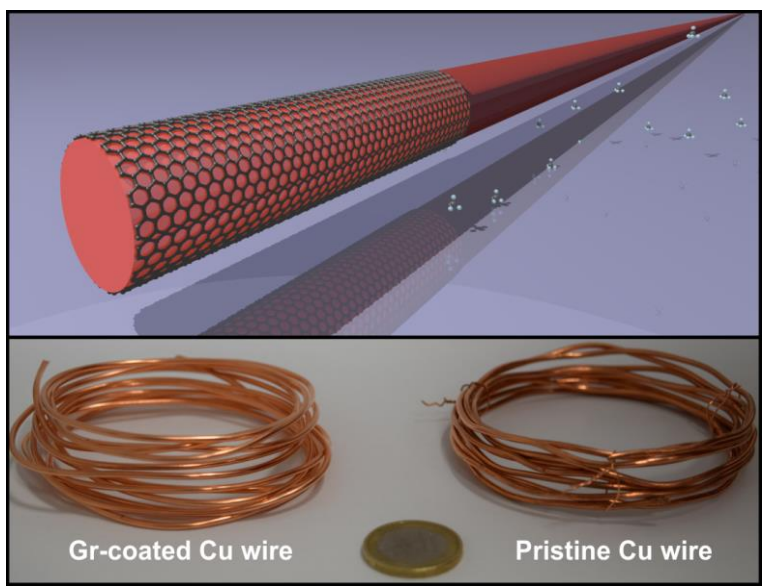